\documentclass[genTeX]{nrc1}

\begin{document}

\title{On The Spectrum of Yang-Mills Theory in 2+1 Dimensions, Analytically}
\author[R.G. Leigh]{R.G. Leigh}
\address{Department of Physics, University of Illinois, Urbana IL 61822, USA. \email{rgleigh@uiuc.edu}}
\author[D. Minic]{D. Minic}
\address{Department of Physics, Virginia Tech, Blacksburg VA 24061, U.S.A.
\email{dminic@vt.edu}}
\author[A. Yelnikov]{A. Yelnikov}
\address{Department of Physics, Virginia Tech, Blacksburg VA 24061, U.S.A.
\email{yelnykov@vt.edu}}

\shortauthor{Leigh, Minic and Yelnykov}

\maketitle
\begin{abstract}
We review our recent work on the glueball spectrum of pure Yang-Mills theory in 2+1 dimensions. The calculations make use of Karabali-Nair corner variables in the Hamiltonian formalism, and involve a determination of the leading form of the ground-state wavefunctional.
\end{abstract}
%

\def\tablefootnote#1{%
\hbox to \textwidth{\hss\vbox{\hsize\captionwidth\footnotesize#1}\hss}} 

\medskip


The understanding of the non-perturbative dynamics of Yang-Mills theory
 is one of the grand problems of theoretical physics. The $2+1$-dimensional theory is expected on many grounds to share the
essential features of its $3+1$ dimensional cousin, such as asymptotic
freedom and confinement, yet is distinguished by the existence of a
dimensionful coupling constant. This simple fact has important consequences.

Here, we describe the determination of the ground state
wave-functional in a specific approximation, the knowledge of which enables us
to determine the mass gap, string tension
and the glueball mass spectrum. The results are in excellent agreement with
the lattice data in the planar limit \cite{teper}. Further details of this work, as well as a full set of references, may be found in \cite{shortpaper,longpaper}. Here we will concentrate on the main ideas and outstanding issues.

We will describe the pure $SU(N)$ Yang-Mills theory by transforming to {\it corner variables} \cite{bars}. One advantage of these variables is that the passage to a lattice regulator is straightforward. This change of variables may be done exactly, the Jacobian of the path integration measure being computable. Given a local coordinatization of space, we introduce the straight Wilson lines
\begin{equation}
M_i(x)=Pexp\left[-\int_{-\infty}^x dx_i A_i\right]
\end{equation}
or equivalently $A_i=-\partial_iM_i\ M_i^{-1}, \forall i.$
Gauge transformations act linearly $M_i\mapsto gM_i$ and local gauge invariant variables may be constructed, $H_{ij}(x)=M_i^{-1}(x)M_j(x)$. These are the (generally constrained) corner variables. In two spatial dimensions, a complex spatial coordinatization may be used, leading to the Karabali-Kim-Nair parameterization\cite{knair,kkn}
\begin{equation}
A=-\partial M\ M^{-1},\ \ \ \ \bar A={M^\dagger}^{-1}\bar\partial M^\dagger,\ \ \ \ H=M^\dagger M
\end{equation}
These variables possess a new local {\it holomorphic invariance}
\begin{equation}
M(z,\bar z)\mapsto M(z,\bar z)h^\dagger(\bar z),\ \ \ \ M^\dagger(z,\bar z)\mapsto h(z)M^\dagger(z,\bar z),\ \ \ \ 
H(z,\bar z)\mapsto h(z)H(z,\bar z)h^\dagger(\bar z)
\end{equation}
which must be preserved. A holomorphic connection for this symmetry is
\begin{equation}
J=\partial H\ H^{-1},\ \ \ \ \ \ J\mapsto h J h^{-1}+\partial h\ h^{-1}
\end{equation}
and the corresponding covariant derivative $D=\partial-J=M^\dagger \nabla {M^\dagger}^{-1}$, with $\nabla=\partial+A$ the usual gauge-covariant derivative.
The path integral measure may be written
\begin{equation}
d\mu[A]\sim d\mu[H] e^{2N S_{WZW}[H]}.
\end{equation}
The Hamiltonian, written as a functional differential operator in $J$'s, was constructed by Karabali-Nair and takes the collective field form
\begin{eqnarray}\label{Hamilt}
{\cal H}_{KN}[J]=T+V=
m \left(\int_x J^a(x) \frac{\delta}{\delta J^a(x)} + 
\int_{x,y}\Omega_{ab}(x,y)
\frac{\delta}{\delta J^a(x)} \frac{\delta}{\delta J^b(y)}\right) +
\frac{2}{g^2} \int_x \bar{\partial} J^a \bar{\partial} J^ a
\end{eqnarray}
where $m$ is the 't Hooft coupling $m=g^2N/2\pi$. In principle, wavefunctionals, expressed as functionals of $J$, may be found by solving the functional Schr\"odinger equation. Note that at weak coupling, the Hamiltonian is dominated by the trailing potential term, while at strong coupling, it is dominated by the kinetic operator. By looking at the action of parity and charge conjugation, it becomes clear that holomorphic invariant wavefunctionals may be constructed from $\bar\partial J = [D,\bar\partial]$ and $\Delta = \{ D,\bar\partial\}$, $\Psi=\Psi[\bar\partial J/m^2, L=\Delta/2m^2]$. As we shall see, in considering the ground state wavefunctional, it appears to be useful to restrict our attention to the gauge and holomorphic invariant form 
\begin{equation}\label{ansatz}
\Psi_0 \simeq exp\left[ -\frac{N}{2\pi m^2} \int tr\ \bar\partial J\ K(L)\bar\partial J+\ldots\right].
\end{equation}
We note that although this is not the most general functional (terms quartic in $\bar\partial J$, etc., are allowed by symmetries), it is highly non-trivial in the sense that the kernel $K(L)$ is allowed to be an arbitrary function, whose exact form will be determined by the Schr\"odinger equation, and its asymptotic features must be consistent with asymptotic freedom {\it and} confinement. We are assuming large $N$ here, which implies a single trace but does not further simplify the exponent. The real importance of large $N$ is in the expectation that interactions amongst gauge invariant states are eliminated and that the states are stable. Note further that the form (\ref{ansatz}) is not equivalent to a simple (covariant) derivative expansion, as all orders in derivatives are included. Since the magnetic field $B=M^\dagger\bar\partial J\ {M^\dagger}^{-1}$, it appears that this may be thought of as a sort of gauge curvature expansion. In a sense, $\bar\partial J$ may be thought of as an adjoint constituent, with glueball states, which are normally identified with vibrating closed strings, modeled as configurations of a pair of constituents connected by strings. Whether or not this picture is borne out, is a matter for experiment to decide. As we detail below, we are able to compute glueball masses which agree remarkably well with the available lattice data. The form (\ref{ansatz}) may not be sufficient for the calculation of other observables.

The first term in the kinetic operator counts the number of $J$'s in a functional; through detailed computations, it appears that the second term in $T$ acts essentially to restore holomorphic invariance. This implies that, defining ${\cal O}_n \equiv tr\ \bar\partial J L^n \bar\partial J$,
\begin{equation}\label{ton}
T {\cal O}_n \simeq (2+n){\cal O}_n + \ldots
\end{equation}
The ellipsis contains mixing with operators containing more factors of $\bar\partial J$ which will not concern us. Eq. (\ref{ton}), crucial to the results that follow, may be explicitly demonstrated for low values of $n$, but has not been firmly established in general. The required calculations rely on a holomorphic invariant regulator and the calculations are tedious; better calculational methods remain to be found. 

Now, given the assumed form of the ground-state wavefunctional (\ref{ansatz}), the Hamiltonian (\ref{Hamilt}) and the result (\ref{ton}), we find that the Schr\"odinger equation takes the form
\begin{equation}
{\cal H}_{KN} \Psi_0 =E_0\Psi_0=\left[\ldots+\int tr\ \bar\partial J\ {\cal R}\ \bar\partial J+\ldots\right]\Psi_0,
\end{equation}
The first ellipsis contains (divergent) terms contributing to the vacuum energy, while the trailing ellipsis contains terms of higher order in $\bar\partial J$. Given the assumed form of the vacuum wavefunctional, this truncation of the Schr\"odinger equation is appropriate and consistent.
The quantity ${\cal R}$ may be found by regarding $K$ as a power series in $L$, and one finds
\begin{equation}
{\cal R} = -K(L)-\frac{L}{2}\frac{d}{dL} K(L) +LK^2(L) +1
\end{equation}
The Schr\"odinger equation requires that we set this to zero, resulting in a differential equation for $K$ of the Riccati type. Through a series of redefinitions, this may be cast as a Bessel equation, and one obtains
\begin{equation}
K(L) = \frac{1}{\sqrt{L}}\frac{ J_2(4\sqrt{L})}{J_1(4\sqrt{L})}
\end{equation}
Although there are other solutions to the differential equation, no other is normalizable in the given path integral measure. We note that although this is a very complicated function (which we now regard as a function of momentum), it has the asymptotics $K(p)\to 2m/p$ as $p\to\infty$ (asymptotic freedom) and $K(p)\to 1$ as $p\to 0$ (confinement). We have thus found that {\it the only normalizable ground state wavefunctional is consistent with both confinement and asymptotic freedom}.
As argued by Karabali and Nair, the infrared limit of our wavefunctional $K$ implies a string tension $\sqrt{\sigma}\simeq \frac{g^2 N}{\sqrt{8\pi}}$ obtained by a dimensional reduction argument. This result agrees precisely with lattice data. However, it is not possible by this line of reasoning to detect different string tensions appropriate to different gauge representations. It is believed that this is consistent, being an artifact of the large $N$ approximation known as Casimir scaling \cite{Greensite} (in particular, this result would be inconsistent at finite $N$). In effect, there is an order of limits problem at large $N$ and large distance.

The ratio of Bessel functions has a rich analytic structure and encodes the mass spectrum of the theory. By Fourier transforming, we find
\begin{equation}
K^{-1}(|x-y|)=-\frac{1}{4\sqrt{2\pi|x-y|}}\sum_{n=1}^\infty (M_n)^{3/2} e^{-M_n|x-y|}
\end{equation}
where $M_n=\gamma_{2,n} m/2$ and $J_2(\gamma_{2,n})=0$. As we will see, this result has direct consequences for the mass spectrum.

To probe the mass spectrum, we consider pair correlation functions of gauge invariant operators with definite spin, parity and charge conjugation quantum numbers. The simplest $0^{++}$ probe operator, for example, is $tr\bar\partial J\bar\partial J$, and we wish to compute
\begin{equation}
\langle tr\bar\partial J\bar\partial J(x)tr\bar\partial J\bar\partial J(y)\rangle
=\int d\mu[H] e^{2NS_{WZW}[H]} \left| \Psi_0\right|^2 tr\bar\partial J\bar\partial J(x) tr\bar\partial J\bar\partial J(y)
\end{equation}
at large spatial separation.
To proceed, we first rewrite the measure as an integral over $J$. Note that if we introduce a variable $\bar J= \bar\partial H H^{-1}$, then we have the `reality condition' $\bar\partial J=[D,\bar J]$.
Given the gauge transformations $\delta A \simeq [\nabla,\delta M M^{-1}],\delta\bar A\simeq [\bar \nabla,M^{-\dagger}\delta M^\dagger]$
, the measure is
\[
\frac{d\mu[A]}{Vol\ G} = det \nabla\bar \nabla\ \frac{d\mu(M,M^\dagger)}{Vol\ G}
=det \nabla\bar \nabla\ d\mu[H]
\]
with $d\mu[H]$ the measure corresponding to $ds^2_{inv}=\int Tr (\delta HH^{-1})^2$. 

We would like to now transform this to the $J$ variables, namely, we would like to find the measure corresponding to the distance $ds^2_J =\int Tr (\delta\bar\partial J)^2$. We first note that $\delta \bar\partial J\simeq [\bar\partial, [D, \delta H H^{-1}]]$. 
So we have
\begin{equation}
 \frac{d\mu[A]}{Vol\ G} = \frac{det \nabla\bar\nabla}{Det \bar\partial D}\frac{d\mu[\bar\partial J]}{Vol\ G_{hol}}
\end{equation}
Since $D=M^\dagger \nabla M^{-\dagger}$ and $\bar\partial=M^\dagger\bar\nabla M^{-\dagger}$, the determinant factor formally cancels exactly. It should be appreciated here though that this is a formal result: in particular, although not apparent, this is a finite volume measure, because the integration is performed over a finite volume region in the complex $\bar\partial J$ space. Ignoring such subtleties we obtain
\begin{equation}
\langle tr\bar\partial J\bar\partial J(x)tr\bar\partial J\bar\partial J(y)\rangle
=\int d\mu[\bar\partial J]  \left| \Psi_0\right|^2 tr\bar\partial J\bar\partial J(x)tr\bar\partial J\bar\partial J(y)
\end{equation}
To the approximation in which we ignore interactions (that is, take $K(L)$ to be a function of momentum), we may thus regard $\bar\partial J$ as a `constituent,' with correlators determined by $K(p)$. Consequently, we find
\begin{equation}
\langle tr\bar\partial J\bar\partial J(x)tr\bar\partial J\bar\partial J(y)\rangle
\simeq
K^{-2}(|x-y|) = \sum_{m,n}\frac{\#}{|x-y|}e^{-(M_n+M_m)|x-y|}
\end{equation}
a form consistent with a single particle $0^{++}$ pole of mass $m_{m,n}=M_m+M_n=(\gamma_{2,m}+\gamma_{2,n}),m/2$. We note that $K^{-1}$ is not the propagator of a physical mode, but it does determine physical propagators. The resulting masses for the $0^{++}$ are collected in the table and compared to the lattice data \cite{teper}. The numbers quoted are in units of the string tension.
\medskip

\begin{tabular}{l|cccc}
\hline\hline
State & Lattice, $N\to\infty$ & Our prediction & Diff, \% \\
\hline
$0^{++}$ & $4.065 \pm 0.055$ & $4.098$ & $0.8$ \\
$0^{++*}$ & $6.18 \pm 0.13$ & $5.407$ & --\\
$0^{++**}$ & $6.18 \pm 0.13$ & $6.716$ & --\\
$0^{++***}$ & $7.99 \pm 0.22$ & $7.994$ & $0.05$\\
$0^{++****}$ & $9.44 \pm 0.38$ & $9.214$ & $2.4$\\
\hline\hline
\end{tabular}
\medskip

We see very good agreement, apart from the second and third states. We note though that the average of these two states coincides with the second lattice state, and we may take this as a prediction that these two states were not resolved in the lattice studies. 

Similar results may be obtained for other states using suitable probe operators and results are equally encouraging. The $0^{--}$ states for example are given in the second table.
\medskip

\begin{tabular}{l|cccc}
\hline\hline
State & Lattice, $N\to\infty$ & Our prediction & Diff,\%\\
\hline
$0^{--}$ & $5.91 \pm 0.25$  & $6.15$ & $4$ \\
$0^{--*}$ & $7.63 \pm 0.37$  & $7.46$ & $2.3$ \\
$0^{--**}$ & $8.96 \pm 0.65$  & $8.73$ & $2.5$ \\
\hline\hline
\end{tabular}
\medskip

The resulting spectrum appears to have an exponentially rising density of states, which may be taken as a manifestation of the effective QCD string.

\end{document}